# Resolving pseudosymmetry in tetragonal ZrO$_2$ using EBSD with a modified dictionary indexing approach

Edward L. Pang, Peter M. Larsen, Christopher A. Schuh

*Department of Materials Science and Engineering, Massachusetts Institute of Technology, 77 Massachusetts Ave, Cambridge, MA 02139, USA*

*Corresponding author. Email address: schuh@mit.edu (C.A. Schuh)

**Abstract**

Resolving pseudosymmetry has long presented a challenge for electron backscatter diffraction (EBSD) and has been notoriously challenging in the case of tetragonal ZrO$_2$ in particular. In this work, a method is proposed to resolve pseudosymmetry by building upon the dictionary indexing method and augmenting it with the application of global optimization to fit accurate pattern centers, clustering of the Hough-indexed orientations to focus the dictionary in orientation space, and interpolation to improve the accuracy of the indexed solution. The proposed method is demonstrated to resolve pseudosymmetry with 100% accuracy in simulated patterns of tetragonal ZrO$_2$, even with high degrees of binning and noise. The method is then used to index an experimental dataset, which confirms its ability to efficiently and accurately resolve pseudosymmetry in these materials. The present method can be applied to resolve pseudosymmetry in a wide range of materials, possibly even some more challenging than tetragonal ZrO$_2$. Source code for this implementation is available online.

# 1. Introduction

In electron backscatter diffraction (EBSD), the incident electron beam in a scanning electron microscope (SEM) is diffracted from the surface of a tilted sample and forms a Kikuchi pattern on the detector, which is then indexed to determine the orientation of the crystal (Venables & Harland, 1973; Wright, 1992; Krieger Lassen, 1994). Traditional indexing techniques, used by all major microscope acquisition software, employ the Hough/Radon transform to identify the band positions on the Kikuchi pattern (Wright, 2000; Krieger Lassen, 1994). The angles between these bands are then compared to a lookup table for the given crystal structure to determine the crystal orientation. For certain materials, however, Hough-based indexing techniques have difficulty identifying the correct orientation. One cause of these difficulties is pseudosymmetry, where the lattice is only slightly distorted from some higher symmetry, such as a tetragonal distortion from cubic symmetry. This gives rise to crystallographically distinct orientations, or variants, which one would like to differentiate but exhibit virtually identical EBSD patterns. A number of materials suffer from pseudosymmetric degeneracy in EBSD characterization, including trigonal $Al_2O_3$ (Nowell & Wright, 2005), γ-TiAl (Zambaldi *et al.*, 2009; Jackson *et al.*, 2018), tetragonal $ZrO_2$ (Nowell & Wright, 2005; Pee *et al.*, 2006; Martin *et al.*, 2012; Ocelík *et al.*, 2017), and many more (Ryde, 2006; Marquardt *et al.*, 2017; De Graef *et al.*, 2019; Nolze *et al.*, 2016).

Some authors have attempted to resolve pseudosymmetry using Hough-based indexing, for example by decreasing the sample-detector distance, increasing the minimum number of bands for indexing, using a higher resolution in the Hough space, and using the band width information (Nowell & Wright, 2005; Ryde, 2006). Zambaldi *et al.* (2009) achieved further improvements by moving to a fit-rank criterion, where the indexed solution is selected based on minimum fit (or mean angular deviation) between the identified band angles and lookup table, as opposed to the default vote-rank criterion used in many software packages, where the solution is selected based on the number of agreements (within a specified tolerance of a few degrees) between the identified band angles and lookup table. However, even with these improvements, Hough-based indexing is still unable to correctly identify pseudosymmetric orientations in γ-TiAl (c/a ~ 1.02) with 100% accuracy (Zambaldi *et al.*, 2009). This is because the small differences in band angles are on the order of the accuracy of band detection by the Hough transform, which implies that Hough transform indexing methods cannot reliably resolve pseudosymmetry in materials with c/a ratio less than ~1.02.

Recently, a number of advanced indexing techniques have emerged that employ full pattern comparison against high-quality dynamically simulated patterns, namely dictionary indexing (Chen *et al.*, 2015; Jackson *et al.*, 2019), spherical indexing (Lenthe *et al.*, 2019*a*), pattern matching (Nolze *et al.*, 2017), and cross-correlation EBSD (Wilkinson *et al.*, 2009; Jackson *et al.*, 2016). These methods exploit the full intensity information in the patterns, rather than just the band positions, and also eliminate the error associated with the band detection step; as such, they are better suited for resolving the subtle differences between pseudosymmetry variants compared to Hough-based indexing (De Graef *et al.*, 2019; Lenthe *et al.*, 2019*b*; Nolze *et al.*, 2016; Jackson *et al.*, 2018). However, many of these techniques are computationally quite expensive, and besides the work by Jackson *et al.* (2018), there have been limited data conclusively demonstrating that the correct variant was selected.

In this work, we develop an alternate intensity-based indexing approach to resolve pseudosymmetry and analyze its ability to correctly distinguish between pseudosymmetry variants. As a test material, we focus on tetragonal $ZrO_2$ (c/a ~ 1.015), a notoriously difficult material in which to resolve pseudosymmetry (Nowell & Wright, 2005; Nolze *et al.*, 2016) and one for which advanced EBSD methods have not yet been applied. To solve this problem, we build upon the dictionary indexing approach (Chen *et al.*, 2015; Jackson *et al.*, 2019) and propose the following modifications: 1) use of our *pcglobal* algorithm to maximize the accuracy of the fitted pattern center (Pang *et al.*, 2020), 2) exploiting information from Hough-based indexing to reduce the size of orientation space needed to explore, and 3) interpolation to improve the accuracy of the final solution. We will first introduce the method in Section 2 and then then demonstrate its ability to correctly distinguish between pseudosymmetry variants on simulated patterns with varying pattern quality in Section 3. Finally, we apply this method on experimental data in Section 4 to demonstrate its accuracy and efficiency in resolving pseudosymmetry in tetragonal $ZrO_2$.

## 2. Proposed method to resolve pseudosymmetry

To illustrate the difficulty of resolving pseudosymmetry in a tetragonal $ZrO_2$-13.5mol% $CeO_2$ material, simulated patterns for three pseudosymmetric orientations are shown in Fig. 1a-c. Patterns were simulated using *EMsoft 4.0* (Jackson *et al.*, 2019) with the parameters given in Table 1. Lattice parameters $a_t$ = 3.631 Å and $c_t$ = 5.230 Å were obtained from Pang *et al.* (2019). Atomic positions and Debye-Waller factors were obtained from ICSD #41579 (Yashima *et al.*, 1995). Inspection of Fig. 1a-c reveals that the three patterns are indistinguishable by eye. Note that the pseudosymmetry is only apparent in the non-primitive double cell (dashed lines in Fig. 1d), with $\sqrt{2}a_t$ = 5.135 Å, which results in c/a = 1.0185. The three variants correspond to aligning the long c-axis with the [0, 0, 1], [1, 1, 0], and [-1, 1, 0] directions in the Cartesian axes shown in Fig. 1d. In this work, all indexings are given with respect to the primitive cell (solid lines in Fig. 1d), unless otherwise noted. In the following subsections, we introduce aspects of the present method used to address pseudosymmetry.

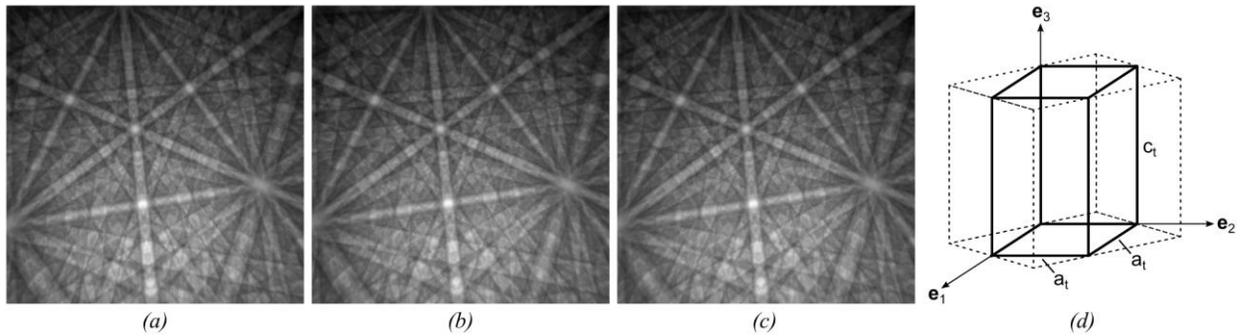

**Fig. 1.** (a-c) Simulated patterns of three pseudosymmetric variants of tetragonal $ZrO_2$-13.5$CeO_2$. (d) Illustration of the primitive unit cell of tetragonal $ZrO_2$ (solid lines) and the non-primitive double cell in which pseudosymmetry is apparent (dashed lines).

**Table 1.** *EMsoft* parameters used in the present study.

| Monte Carlo and master pattern simulation parameters | | Pattern simulation parameters | |
|---|---|---|---|
| Total number of incident electrons | $2\times10^9$ | Camera elevation | 5° |
| Specimen tilt angle | 70° | Incident beam current | 150 nA |
| Incident beam energy | 25 keV | Beam dwell time | 100 µs |
| Minimum BSE exit energy | 10 keV | Gamma value | 0.33 |
| Energy bin size | 1 keV | Detector size | 480×480 px |
| Maximum exit depth | 100 nm | Detector pixel size | 50 µm |
| Depth step size | 1 nm | Bit depth | 8 bit |
| Smallest d-spacing | 0.05 nm | | |
| Master pattern size | 1001×1001 px | | |

*2.1. Fitting the pattern center*

An accurate pattern center is necessary to obtain accurate orientation data using intensity-based indexing methods (Ram *et al.*, 2017; Tanaka & Wilkinson, 2019; Pang *et al.*, 2020). In pseudosymmetric materials, this is even more important since crystallographically distinct orientations can give rise to virtually identical patterns, as previously reported by (Zambaldi *et al.*, 2009; Jackson *et al.*, 2018; De Graef *et al.*, 2019) and seen in Fig. 1. To investigate the effect of pseudosymmetry on our ability to fit accurate pattern centers, we performed tests on simulated patterns of tetragonal $ZrO_2$-13.5$CeO_2$, where the true pseudosymmetry variant and pattern center of each pattern are known. Ten patterns were simulated with random orientations obtained using *MTEX 5.1.1* and a true pattern center ($X^*$, $Y^*$, $Z^*$) of (0 px, 80 px, 15000 μm). To explore the effect of pattern quality, simulated patterns were subjected to various levels of artificial noise and binning. Gaussian noise was added using the `imnoise` function in *MATLAB*. Variances of 0.03, 0.15, and 0.75 (in fraction of the intensity range) were used to generate images with peak signal-to-noise ratios (PSNR) of approximately 15.6, 10.3, and 7.4 dB, respectively. The patterns were then binned by a factor of 1, 2, 4, or 8, giving final image sizes of 480×480, 240×240, 120×120, and 60×60, respectively. Representative images are shown in Fig. 2.

Pattern center fittings were performed with our recent *pcglobal* package (Pang *et al.*, 2020). This method uses *EMsoft* to generate dynamically simulated patterns with varying pattern center and orientation to compare with the experimental pattern, in search of a solution that maximizes the similarity between the simulated and experimental patterns, as measured by the normalized dot product (NDP). The NDP can vary from 0 to 1, with higher values representing better matching between the images. The *pcglobal* package employs the SNOBFIT global optimization algorithm (Huyer & Neumaier, 2008) along with transformation and scaling of variables to a more uniform fitting space, which maximizes the accuracy of the fitted pattern center. We started all fittings from a perturbed pattern center of (-2.3664 px, 78.8267 px, 15135.51 μm), which corresponds to errors of (-0.49%, -0.24%, +0.56%) in percent detector width. The starting orientation for each pattern was perturbed by adding a random value within ±0.5° (from a uniform distribution) to the true Euler angles. Trust radii used for the SNOBFIT algorithm were 10 px in $X^*$ and $Y^*$, 500 μm in $Z^*$, and 1.5° in orientation. All other parameters were identical to those used in (Pang *et al.*, 2020).

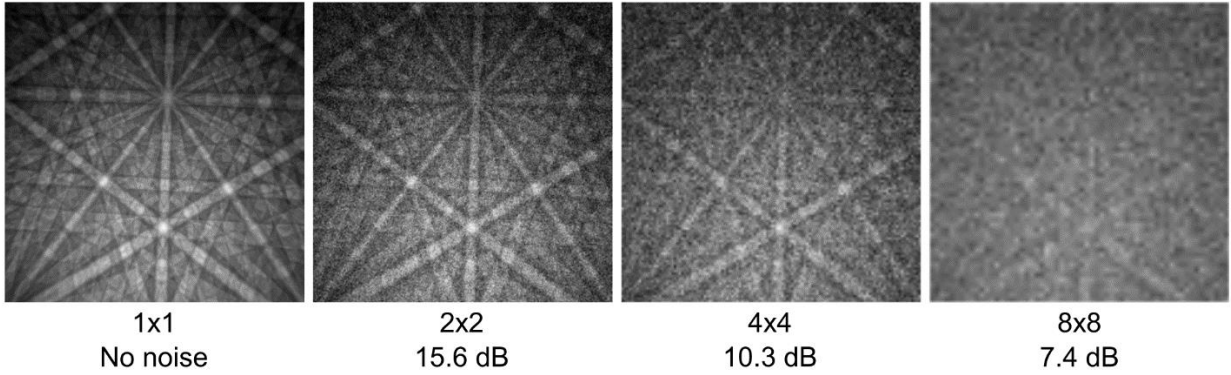

**Fig. 2.** Simulated patterns of tetragonal $ZrO_2$-13.5$CeO_2$ with various levels of binning and noise.

In pseudosymmetric materials such as tetragonal $ZrO_2$, it is unknown a priori which pseudosymmetric orientation we should fit to, and these orientations lie in different regions of orientation space that are not explored during the local pattern center fitting process. Thus, the pattern center fitting was performed for the original variant as well as the two additional pseudosymmetry variants, whose orientations were obtained by the following rotations:

90° about [110]

90° about [1$\bar{1}$0].  (1)

Results from the pattern center fitting are shown in Fig. 3 for varying levels of binning and noise. For all 10 patterns tested at each binning and noise level, the correct variant was found to have the highest NDP, even for the patterns with 8×8 binning and a PSNR of 7.4 dB where the pattern is barely recognizable to the human eye (Fig. 2). The confidence index (CI), which we take to be ΔNDP between the variants with the highest and second-highest NDP, is plotted in Fig. 3a. The CI continually decreases with additional binning and noise, but CI values for the 1×1, 2×2, and 4×4 cases are orders of magnitude higher than the resolution of the SNOBFIT algorithm in finding the global maximum, which we previously showed to be less than 0.0001 (Pang *et al.*, 2020). This gives confidence that the *pcglobal* package can distinguish the correct pseudosymmetry variant in tetragonal $ZrO_2$-13.5$CeO_2$ for all but the highest levels of binning and noise.

Fig. 3b-d shows the range of pattern center errors obtained by fitting to a single pattern, both for the actual variant (left, blue) and the incorrect variant with the second-highest NDP (right, blue). While small errors are obtained by fitting to the correct variant, extremely large errors of up to 2% detector width can be obtained by fitting to the incorrect variant. Jackson *et al.* (2018) have found that the pattern center needs to be known to a much higher degree of accuracy, within ~0.4% detector width, to successfully resolve pseudosymmetry in γ-TiAl with a similar c/a ratio of 1.02 by direct pattern comparison using a similar cross-correlation metric. Thus, it is crucial that the correct variant be selected during the pattern fitting step; otherwise, the ability of the subsequent indexing to correctly resolve pseudosymmetry will be compromised. We therefore recommend careful optimization by using a global search algorithm such as SNOBFIT and properly scaled variables, as implemented in the *pcglobal* package, to prevent cases where the incorrect variant is selected because of optimization errors. In addition, we also recommend using average results from multiple patterns (typically 10) from differing regions or grains to obtain the best possible pattern center estimate.

We averaged the results from the 10 patterns fitted to the correct variant in Fig. 3, and the resulting errors are shown in Table 2. These results are in good agreement with our previous report for non-pseudosymmetric α-Fe (Pang *et al.*, 2020), which is validating given that the correct variant was selected in each case here. In the case of $ZrO_2$-13.5$CeO_2$, with c/a = 1.0185, it appears that using the *pcglobal* package to fit to each of the three pseudosymmetry variants and then selecting the one with the highest NDP is an effective strategy for fitting the pattern center, even with noisy binned patterns. However, this surely will be more difficult for materials with lower c/a ratios; this is currently an ongoing area of investigation, and these results will be presented in a forthcoming publication.

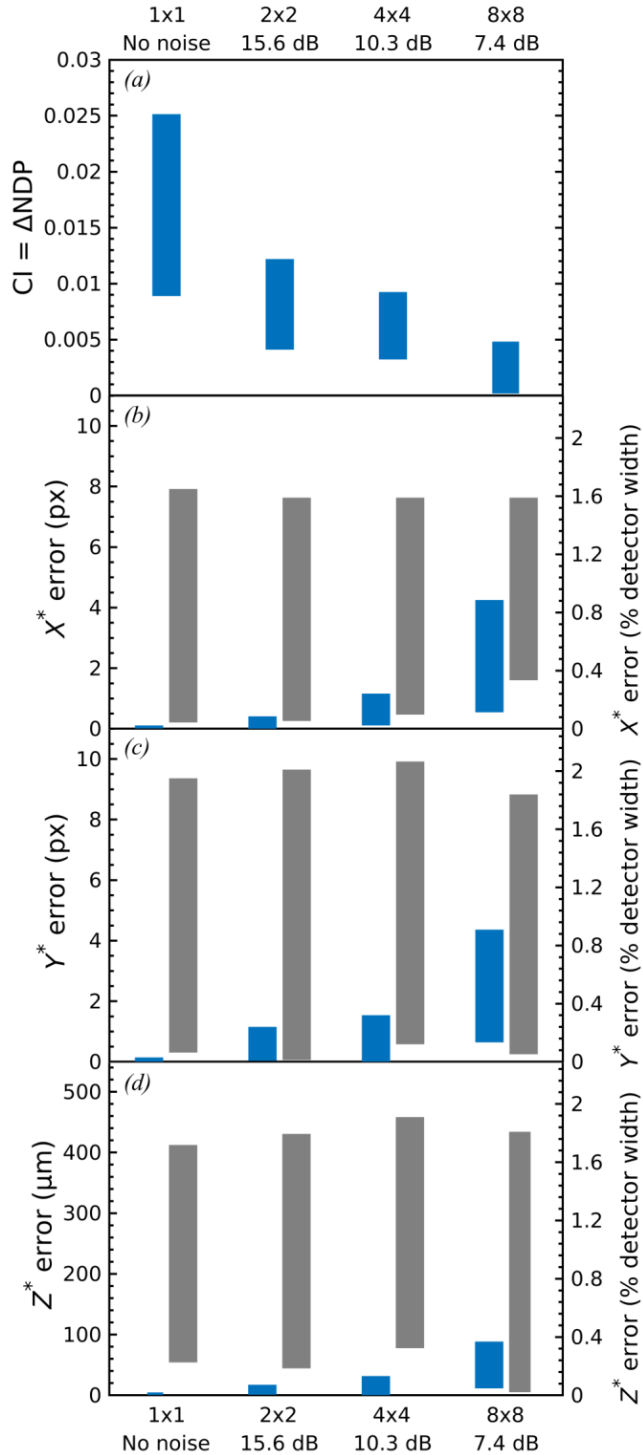

**Fig. 3.** Pattern center fitting errors for 10 simulated patterns of $ZrO_2$-13.5$CeO_2$ with varying levels of binning and noise using the *pcglobal* package. (a) Range of $\Delta$NDP, or confidence index (CI), observed over the 10 separate patterns. For all levels of binning and noise, the correct variant was selected for all 10 patterns. (b-d) Range of absolute pattern center errors observed over the 10 separate patterns. The left (blue) bar of each pair represents the variant with the highest NDP, and the right (gray) bar corresponds to the variant with the second highest NDP.

**Table 2.** Mean and standard deviation of the pattern center errors found using the *pcglobal* package on 10 simulated patterns of $ZrO_2$-13.5$CeO_2$ with varying levels of binning and noise.

|  | X* | Y* | Z* |
|---|---|---|---|
| 1×1 No noise | -0.033 ± 0.055 px (-0.0069 ± 0.011%) | -0.047 ± 0.094 px (-0.0098 ± 0.020%) | 1.7 ± 2.0 μm (0.0072 ± 0.0083%) |
| 2×2 15.6 dB | -0.059 ± 0.21 px (-0.012 ± 0.043%) | 0.11 ± 0.39 px (0.023 ± 0.081%) | 5.1 ± 5.7 μm (0.021 ± 0.024%) |
| 4×4 10.3 dB | 0.15 ± 0.83 px (0.032 ± 0.17%) | -0.33 ± 0.79 px (0.068 ± 0.16%) | 10.4 ± 17.1 μm (0.043 ± 0.071%) |
| 8×8 7.4 dB | 0.70 ± 2.41 px (0.15 ± 0.50%) | -0.49 ± 2.48 px (-0.10 ± 0.52%) | 15.7 ± 40.0 μm (0.065 ± 0.17%) |

*2.2. Reduced orientation dictionary*

Dictionary indexing is not ideal for resolving pseudosymmetry because of its discrete nature. Traditionally, dictionary indexing involves generating a discrete list of all possible orientations within the fundamental zone for the given symmetry (the "dictionary"), computing the NDP between simulated patterns of each of these orientations against each experimental pattern, and selecting the orientation giving the highest NDP as the indexed orientation for each pattern (Chen *et al.*, 2015; Jackson *et al.*, 2019). For computational reasons, dictionaries typically have grid spacings of ~1.5°, which is just fine enough to ensure that at least one grid point falls in the NDP peak (Singh & De Graef, 2016). If there is only one significant peak in the entire space, such as for non-pseudosymmetric materials, then this is satisfactory. However, if there are multiple peaks of similar height in NDP space, as for pseudosymmetric materials, then the grid needs to be extremely fine to identify the tallest peak, which is computationally impractical. For example, even using a dictionary with an angular resolution of 0.5° (cubochoric N = 285), which took 76 hours to run on our 8-core workstation (see Section 4.1 for details), we could not fully resolve pseudosymmetry in an experimental dataset of $ZrO_2$-13.5$CeO_2$. A more efficient indexing strategy is therefore desirable for such subtle situations, and we propose to limit the dictionary search space to achieve such efficiency, as explained below.

The vast majority of dictionary orientations are not near any orientations of interest in a given dataset. For example, the experimental dataset that we will examine later in this paper contains 12 grains with tetragonal symmetry (point group 4/mmm) and only covers 0.15% of orientation space, assuming non-overlapping regions of orientation space each spanning 3°. Thus, almost all of the time taken for dictionary indexing is evaluating orientations that are irrelevant for the dataset (Foden *et al.*, 2019; Lenthe *et al.*, 2019*a*). However, the Hough-indexed data does inform what regions of orientation space to focus on. Based on this information, a dictionary focused around these orientation clusters can be constructed to improve the computational efficiency.

We form clusters by grouping orientations from the Hough-indexed data that are separated by a disorientation angle below a specified threshold (we use 1.2°). During this stage, we can also optionally ignore clusters that contain less than a certain number of points or have a fit metric (mean angular deviation) above a certain threshold to filter out misindexings from the Hough data that will slow down the analysis. We note similarities with the multivariate statistical analysis (MSA) approach (Brewer *et al.*, 2008; Wilkinson *et al.*, 2019), but in those studies the analysis was performed on the patterns themselves, as opposed to their orientations, to reduce the number of patterns to index. Once clusters have been identified, we also add the pseudosymmetric orientations for each cluster to this list by applying the rotations in Eq. (1).

We then create fine orientation grids around each of these cluster orientations that will become part of the high-resolution dictionary. Orientation grids were created by starting with a simple cubic grid in Rodrigues space (Engler & Randle, 2010) centered about [0, 0, 0] with each of the three components of the Rodrigues vector [$R_x$, $R_y$, $R_z$] having a value in the range

$$\left[-\tan\left(\frac{\alpha}{2}\right) \quad ... \quad \tan\left(\frac{\alpha}{2}\right)\right] \text{ in steps of } \frac{1}{N}\tan\left(\frac{\alpha}{2}\right). \qquad (2)$$

This gives a grid of size $(2N+1)^3$ that checks misorientations up to ±α in each principal direction from the grid center. The Rodrigues vectors of each grid point are then rotated so that the grid is centered about the orientation of interest. We choose to generate the grid as misorientations in Rodrigues space because for small misorientation angles, this space is essentially uniform and contains no singularities, unlike Euler angles. In this manner, we obtain a focused dictionary with much finer angular resolution than the standard 1.5° of dictionary indexing, but without an increase in dictionary size and associated computational cost during indexing.

*2.3. Interpolation to find peak maxima*

One major drawback to the method outlined thus far is that the indexed orientation is still discrete. This problem has been solved for dictionary indexing by performing an orientation optimization on each map point after the initial dictionary indexing run (Singh *et al.*, 2017; De Graef *et al.*, 2019). However, this approach is rather computationally intensive, and this refinement step can take even longer than the initial dictionary indexing run. Instead, we choose to use cubic interpolation to find the maximum of the NDP peak. In addition to the benefit of speed, there may be benefits in accuracy as well, as it is easy for downhill search algorithms to get stuck in local minima presented by the noisy optimization landscape (Rios & Sahinidis, 2013; Tanaka & Wilkinson, 2019; Pang *et al.*, 2020). Since we have already sampled the NDP on a grid from our refined dictionary, we can easily perform cubic interpolation to estimate the NDP and misorientation of the peak tip. We then convert the misorientation in Rodrigues space, Eq. (2), back to an orientation.

To validate this approach, we performed interpolation using various grid sizes and compared the interpolated peak with that found by the SNOBFIT global optimization algorithm. Fig. 4a illustrates the parameters that were varied, namely the half-width *w* and the number of grid points (in this example 7×7×7, which corresponds to N = 3 in Eq. (2)). The centers of the grids were shifted away from the true orientation by adding *w*/5 to each Euler angle to ensure that a grid point was not located at the peak maximum. The resulting errors in NDP, $e_{NDP}$, and orientation, $e_\theta$, (illustrated in Fig. 4b) are displayed in Fig. 4c-j for varying levels of binning and noise.

We find that different trends emerge at low *w* values compared to high *w* values. At low *w* values, both $e_{NDP}$ and $e_\theta$ increase as the pattern quality is degraded, as one would intuitively expect. A comparison of the actual and interpolated linescans reveals that in all cases, the large-scale shape of the true NDP function is well predicted by the interpolated function, but the increased noise at poorer pattern quality limits the accuracy of the interpolation. At high *w* values, the average NDP error counterintuitively decreases as the pattern quality is degraded. For example, $e_{NDP}$ for the 5×5×5 grid at $w = 2.5°$ is equal to 0.070, 0.036, 0.026, and 0.012 in order of decreasing pattern quality. In this case, the spacing between grid points is much greater than the scale of the noise, so noise does not limit the interpolation accuracy. Closer analysis reveals that the introduction of binning and noise to the patterns decreases the height and increases the width of the NDP peaks, which leads to a lower curvature near the peak tip. Thus, cubic interpolation can more accurately predict the peak height. This effect is less pronounced for $e_\theta$, which only slightly decreases with degrading pattern quality, since prediction of the peak location by interpolation is less sensitive to the curvature near the peak tip. We will focus on the high-$w$ regime, as relatively wide grids are needed when the grid centers are chosen from Hough-indexed data, which have a reported precision less than 0.5° (Demirel *et al.*, 2000; Brough *et al.*, 2006; Wright & Nowell, 2008) but can have absolute errors up to 2° (Humphreys, 1999, 2001), to ensure that all peaks are located within the span of each grid.

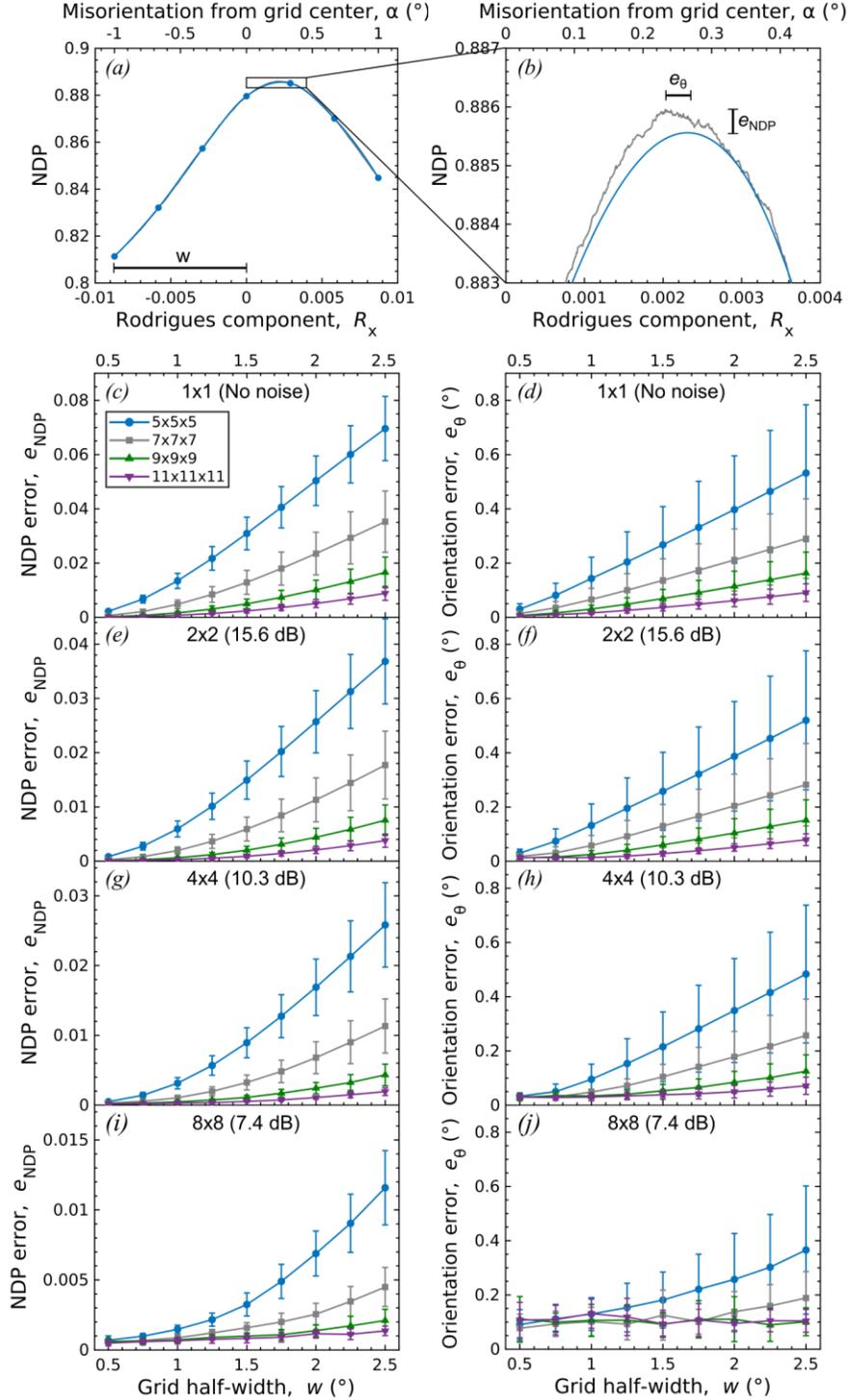

**Fig. 4.** Interpolation accuracy as a function of the interpolation grid size and half-width for simulated patterns of $ZrO_2$-$13.5CeO_2$ with varying binning and noise. Results are averaged over 10 patterns, and error bars represent one standard deviation. (a-b) Illustration of the interpolation parameters $w$, $e_{NDP}$, and $e_\theta$. The smooth blue curve is the interpolated NDP function, and the jagged gray curve is the actual NDP function. (c,e,g,i) NDP error from interpolation. Note the varying y-axis scales. (d,f,h,j) Orientation error from interpolation.

*2.4. Summary of the method*

At this point, we summarize the entire method as in Fig. 5 or in the following steps:

1) Estimate the pattern center by averaging the results from multiple patterns (we use 10), ideally from different grains, using the *pcglobal* package.
2) Perform a clustering analysis with a specified radius (we use 1.2°), on the Hough-indexed orientations (reduced to the fundamental zone). Clusters containing less than a certain number of points or having a fit metric above a certain threshold can be optionally ignored to filter out misindexings from the Hough data.
3) Construct a fine-resolution orientation grid (we use 0.5°) around the center of each cluster identified in step 2 and all pseudosymmetric orientations.
4) Compile all of the orientation grids into a single focused dictionary.
5) Perform a dictionary indexing run using this focused dictionary.
6) Loop through each map point and perform interpolation. This involves the following steps:
    a) Identify which grid center is closest (lowest misorientation) to the indexed orientation, and extract the NDP versus orientation data for that grid and its pseudosymmetric grids.

b) Perform cubic interpolation to find the peak location (orientation) and peak height (NDP) for each pseudosymmetry variant.

c) Select the variant with the highest NDP. Report ΔNDP between the best variant and second-best variant as a confidence index.

7) Repeat steps 2-6 on additional phases, if necessary. Once all phases have been indexed, select the phase giving the highest NDP for each map point.

In the rest of the paper, we evaluate the ability of this method to correctly resolve pseudosymmetry in both simulated and experimental datasets of $ZrO_2$-13.5$CeO_2$.

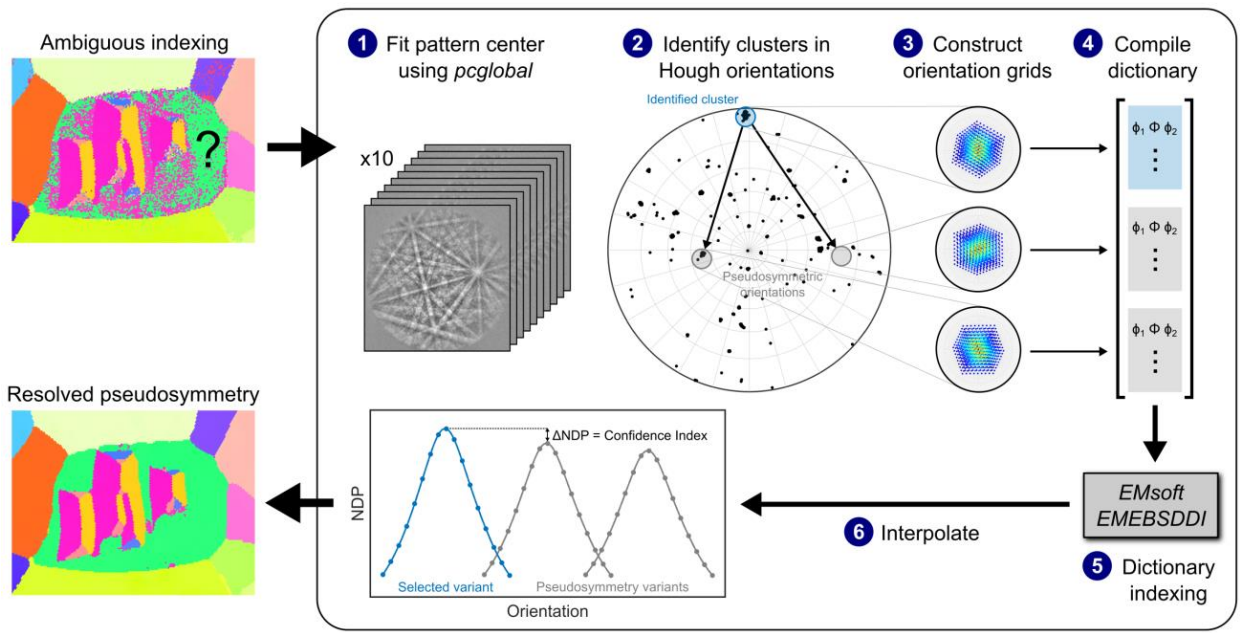

**Fig. 5.** Schematic of the present modified dictionary indexing method used to resolve pseudosymmetry.

## 3. Resolving pseudosymmetry in a simulated $ZrO_2$ dataset

First, we need to determine the interpolation grid parameters necessary to distinguish between pseudosymmetry variants in zirconia during indexing. To estimate ΔNDP between variants, we performed orientation optimization on the same 10 patterns used in Section 2.1 for the pattern center fitting, but here we fixed the pattern center to the values in Table 2 as is the case during indexing. Each pattern was subjected to an orientation optimization using the SNOBFIT algorithm. The ΔNDP values

obtained were 0.030, 0.015, 0.010, and 0.004 in the case of 1×1 (no noise), 2×2 (15.6 dB), 4×4 (10.3 dB), and 8×8 (7.4 dB) patterns, respectively. Note that these values are higher than the values shown in Fig. 3a since here the pattern center was not allowed to vary during the optimization. Comparing with the NDP interpolation accuracy data from Fig. 4, we estimate that a 9×9×9 grid with $w = 2°$, which is expected to give NDP errors of 0.010, 0.0044, 0.0024, and 0.0014, should be sufficient to correctly distinguish between $ZrO_2$ pseudosymmetry variants since these errors are less than half of the minimum ΔNDP between variants.

To test the accuracy of our method outlined in Fig. 5, we simulated 100 patterns of $ZrO_2$-13.5$CeO_2$ with random orientations in the same manner as described in Section 2.1. The starting orientation was perturbed by randomly rotating to one of the pseudosymmetric orientations with equal probability and then adding a random amount between ±0.5˚ (drawn from a uniform distribution) to the resulting Euler angles to simulate the error from Hough-indexed data. For each level of binning/noise, we used the pattern center values from Table 2 to build into the test results our ability to estimate the correct pattern center. We then indexed this dataset for varying levels of binning and noise (representative patterns shown in Fig. 2) using a 9×9×9 interpolation grid with $w = 2°$.

Fig. 6a shows the disorientation between the indexed orientation and the true orientation, which shows no values near 90° and thus demonstrates that all 100 patterns were indexed to the correct pseudosymmetry variant for all levels of binning and noise. We find that orientation accuracy initially is not strongly affected by pattern quality, as the peak of the disorientation distribution appears to remain approximately constant at ~0.08° for the 1×1 (no noise), 2×2 (15.6 dB), and 4×4 (10.3 dB) cases. It is not until the pattern quality is further degraded to 8×8 (7.4 dB) that the peak error increases significantly to ~0.2°. This trend differs from that shown by the interpolation error results in Fig. 4, because here the error from interpolation is convoluted with the increasing orientation error obtained at larger pattern center errors (Ram *et al.*, 2017; Tanaka & Wilkinson, 2019; Pang *et al.*, 2020). Confidence index (CI) values are shown in Fig. 6b, which reveals that CI continually decreases with degrading pattern quality. Minimum CI values of 0.0160, 0.0092, 0.0074, and 0.0024 were obtained in order of degrading pattern quality, which are consistent with our estimates above once the interpolation error is also taken into account.

Thus, we conclude that the present method correctly resolves pseudosymmetry in tetragonal $ZrO_2$-13.5$CeO_2$ (c/a = 1.0185) over a wide range of orientations and pattern qualities. For highest confidence in resolving pseudosymmetry, large high-quality patterns should be used. However, it appears that typical mapping conditions with 2×2, 4×4, or even 8×8 binning may be sufficient to correctly resolve pseudosymmetry in these materials using the present method. Although, surely this performance represents an upper bound, as experimental data possesses additional uncertainties in strain state, lattice parameters, and optical distortions, among other sources of error.

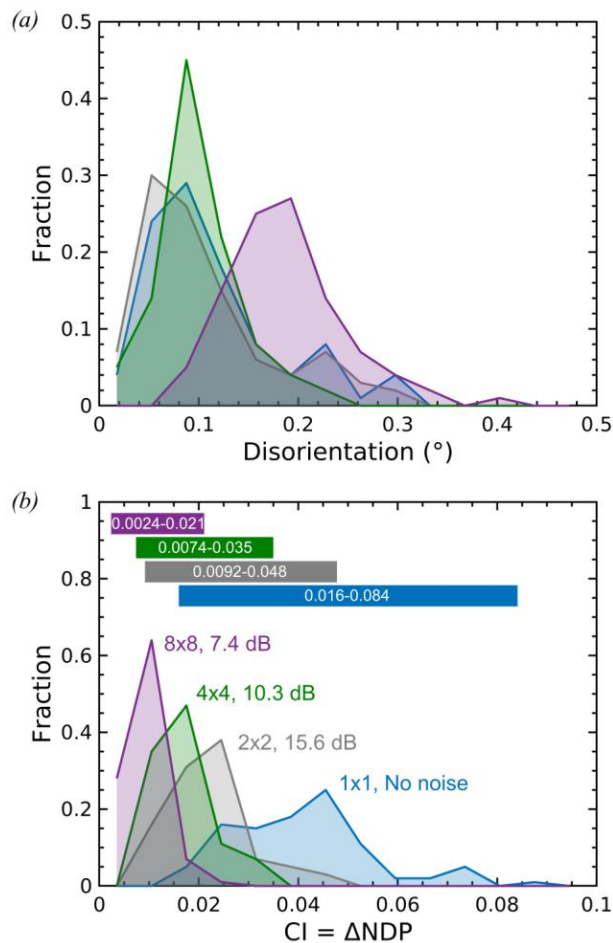

**Fig. 6.** Performance of the present indexing method on a simulated dataset containing 100 random orientations of $ZrO_2$-13.5$CeO_2$ for varying levels of binning and noise. (a) Distribution of disorientations between the indexed and true orientations. See part b for labels. (b) Distribution of CI values. Horizontal bars at the top of the figure denote the full range of CI values observed.

## 4. Resolving pseudosymmetry in an experimental ZrO$_2$ dataset

To validate the performance of our method beyond simulated data, we used the same method to index an experimental dataset of a ZrO$_2$-13.5mol% CeO$_2$ material, which was previously reported by Pang *et al.* (2019). This dataset is interesting in that it contains both pseudosymmetric tetragonal phase as well as non-pseudosymmetric monoclinic phase in a specific lattice correspondence, which allows us to verify numerous aspects of the EBSD indexing.

*4.1. Methods*

EBSD was performed in a Zeiss Merlin field emission SEM (FE-SEM) equipped with an EDAX Hikari XP detector and *OIM Data Collection 7* acquisition software. Samples were prepared by conventional mechanical polishing techniques and then cooled to -253°C and held for 1 h in an Oxford PheniX cryostat under vacuum to induce partial transformation to monoclinic phase from the parent tetragonal phase. Samples were then coated with ~5 nm of carbon. Patterns were acquired using an accelerating voltage of 25 kV, beam current of 10.7 nA, sample tilt of 70°, 2×2 camera binning (resulting in patterns of size 240×240 px), gain of 4.85, and exposure of 47.03 ms, which resulted in an acquisition rate of 21.2 fps. Patterns underwent background subtraction, dynamic background correction, and intensity histogram normalization in the vendor software.

Hough-based indexing was performed in the vendor software using a binned pattern size of 120 px, theta step size of 0.5°, peak count 7 to 12, and classic low resolution Hough with a 9×9 convolution mask. For indexing, an interplanar angle tolerance of 1° was used in conjunction with fit ranking, which has been demonstrated to be more effective in resolving pseudosymmetry than the default triplet vote ranking (Zambaldi *et al.*, 2009). Data was indexed using a pattern center of ($X^*$, $Y^*$, $Z^*$) = (0.49951, 0.68292, 0.66096), given in fraction detector width in reference to the EDAX/TSL coordinate system. *EMsoft 4.0* (Jackson *et al.*, 2019) was used for dictionary indexing. The same lattice parameters were used for the Hough indexing and dictionary indexing, which were previously measured on the same sample by X-ray diffraction (Pang *et al.*, 2019): $a_t$ = 3.631 Å and $c_t$ = 5.230 Å for tetragonal phase, the same as for the simulated tests in Sections 2 and 3, and $a_m$ = 5.216 Å, $b_m$ = 5.225 Å, $c_m$ = 5.397 Å, and $β_m$ = 98.797° for monoclinic phase. Atomic positions and Debye-Waller factors were obtained from ICSD #41579 (Yashima *et al.*, 1995) and #183856 (Jimenez *et al.*, 2011) for the tetragonal and monoclinic phases, respectively. Calculations were run on a Linux workstation, with an 8-core Intel i7-7820X CPU and an NVIDIA GeForce GTX 1080 GPU. Maps were plotted using *OIM Analysis 7* software.

*4.2. Resolving pseudosymmetry*

Results from the Hough-based indexing are shown in Fig. 7a. Some grains of tetragonal phase (note the phase map in Fig. 7c) show a clear single orientation whereas some grains show a "checkerboard" pattern, where the indexing routine cannot decide on a single pseudosymmetry variant. This is consistent with previous reports of Hough-based indexing on pseudosymmetric materials (Ryde, 2006; Wright, 2006; Nolze *et al.*, 2016; Ocelík *et al.*, 2017). In addition, the reported confidence index for all points indexed as tetragonal phase is low, with a majority below the accepted "good" value of 0.1 (Wright, 2000). We took extra care to optimize the indexing parameters for resolving pseudosymmetry, yet it is clear that Hough-based indexing cannot resolve pseudosymmetry for certain orientations.

We then re-indexed the same dataset using the method outlined in this paper. First, we fit the pattern center using the *pcglobal* package to 10 patterns of tetragonal phase, one from each of the 10 largest grains and away from grain boundaries and second phases. We note that these were patterns from the dataset itself rather than separate unbinned high-quality patterns. One such pattern is shown in Fig. 7f. This resulted in a pattern center (and standard deviation) of

$X^* = 1.19 \pm 0.78$ px ($0.50248 \pm 0.00163$ in fraction detector width)

$Y^* = 93.70 \pm 0.90$ px ($0.69521 \pm 0.00188$)

$Z^* = 15696 \pm 41$ µm ($0.65400 \pm 0.00171$). (3)

The minimum CI was 0.0042, which is similar to that obtained on comparable simulated patterns with 2×2 binning and small amounts of noise (Fig. 3a). To check the accuracy of this pattern center fit, we also fit the pattern center to 10 patterns of the non-pseudosymmetric monoclinic phase, which gave

$X^* = 0.94 \pm 0.88$ px ($0.50196 \pm 0.00183$)

$Y^* = 92.83 \pm 0.95$ px ($0.69340 \pm 0.00198$)

$$Z^* = 15733 \pm 18 \text{ μm } (0.655542 \pm 0.000750). \tag{4}$$

These two pattern centers are in good agreement, with no statistical significance between them. In addition, the standard deviation of $X^*$ and $Y^*$ fitted to the tetragonal phase is no larger than that for the monoclinic phase. The spread of $Z^*$ is a bit higher for the tetragonal phase than for the monoclinic phase but is still rather small. Considering that fitting to the incorrect variant can give rise to extremely large errors up to 2% detector width (Fig. 3), this gives confidence that the pattern center can be accurately fit on experimental patterns of pseudosymmetric materials by averaging results over 10 patterns using the *pcglobal* package. However, when possible, it is still preferable to fit the pattern center on a non-pseudosymmetric phase, for which there is no additional uncertainty. As such, we proceeded to index the dataset using the pattern center in Eq. (4) fit on the monoclinic phase.

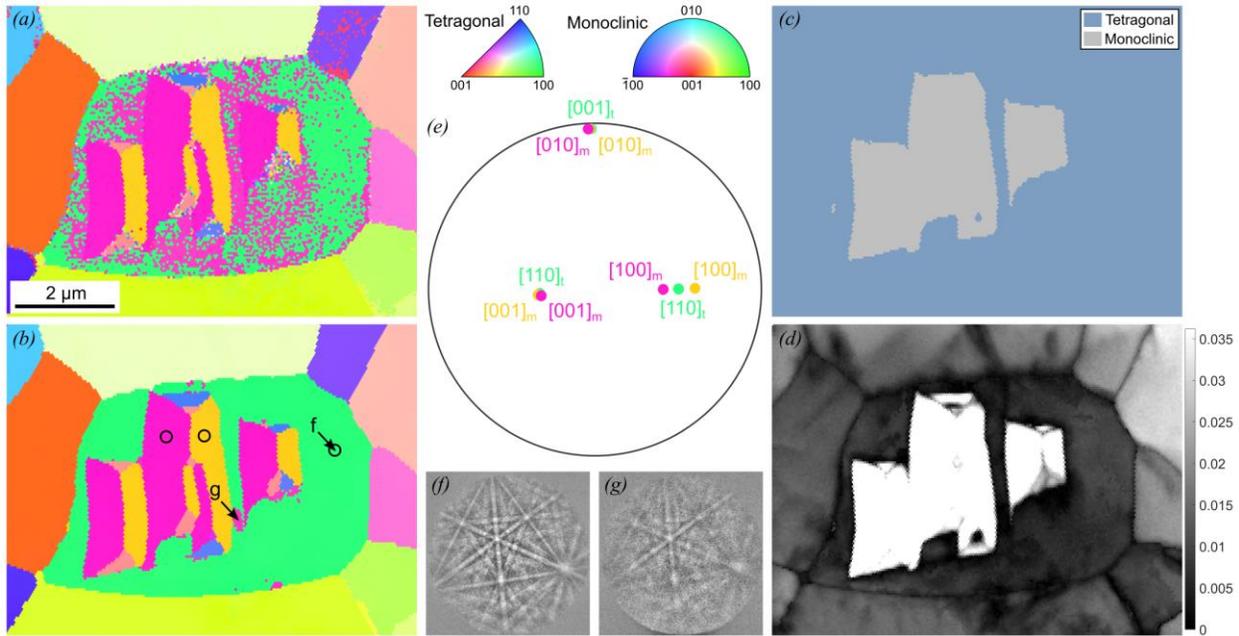

**Fig. 7.** EBSD data from a two-phase $ZrO_2$-13.5$CeO_2$ ceramic. (a) Inverse pole figure (IPF) map of the Hough-indexed data. (b) IPF map, (c) phase map, and (d) CI map obtained using the indexing method presented in this work. (e) Pole figure showing the lattice correspondence between the tetragonal and monoclinic phases. Orientation data taken from the circled points in part b, color-coded to match the IPF map. (f) A representative pattern of tetragonal phase from this dataset, from the labeled location in part b. (g) A lower quality pattern, from the labeled location in part b.

For the initial clustering of Hough orientations, we chose to limit the cluster radius to 1.2°. We only considered points that were indexed with a fit less than 0.5° for tetragonal phase and 1° for monoclinic phase, and we also eliminated clusters containing less than 5 points. Both of these steps were taken to eliminate random misindexings that would needlessly increase the size of the dictionary and slow down the analysis. For the interpolation grid, we used a 9×9×9 grid with half-width $w = 2°$, the same as used throughout this paper. This resulted in 15 retained clusters for the tetragonal phase, which led to a dictionary size of 32,805 orientations after adding in pseudosymmetry variants. This took 45 min to run on our workstation, compared to 2.7 hours for a standard dictionary indexing run with 1.5° grid spacing (cubochoric N = 90, dictionary size of 730,112), which is unable to resolve pseudosymmetry in this material. Additional refinement to resolve pseudosymmetry using the *EMFitOrientation* program within *EMsoft* is estimated to take an additional 9.4 hours on this workstation. For the monoclinic phase, 18 clusters were retained, which led to a dictionary size of 13,122 orientations. This took 19 minutes to run on our workstation, compared to 9.6 hours for standard dictionary indexing (N = 90, dictionary size of 2,916,352).

The inverse pole figure (IPF) map of this indexed data is shown in Fig. 7b, which eliminates the "checkerboard" pattern seen in the Hough-indexed data and selects a single variant in each tetragonal grain. The phase map is shown in Fig. 7c, which was obtained by stitching together the separate tetragonal and monoclinic phase indexing runs, selecting for each map point the phase with the highest NDP. A CI value was assigned to each point based on ΔNDP between the highest peak and next highest peak, which could be from a different phase or pseudosymmetry variant. The CI map is shown in Fig. 7d, which reveals very high confidence for the monoclinic phase compared to the pseudosymmetric tetragonal phase, as expected. The grain with the lowest CI is the central grain, with an average CI value of ~0.007. The grain with the highest CI is the peach-colored grain in the top-right, with an average CI value of ~0.030. This observed range of CI values is in excellent agreement with that observed in our simulated dataset with 2×2 binned patterns, 0.0074-0.035.

To validate our indexing, we analyzed the lattice correspondence between the tetragonal and monoclinic phases. In $CeO_2$-doped $ZrO_2$, the preferred lattice correspondence is such that $[001]_t$ || $[010]_m$, which results in low energy interfaces between the two phases (Kelly & Francis Rose, 2002;

Pang *et al.*, 2019). In Fig. 7b, the plates of monoclinic phase are elongated in the vertical direction, which suggests that the long vertical interfaces between tetragonal and monoclinic phase are low in energy. Fig. 7e shows a pole figure depicting the orientations of the relevant crystal axes for the regions circled in Fig. 7b, which reveals that the large monoclinic plates are indeed oriented such that $[010]_m$ is aligned with $[001]_t$, and not $[110]_t$. This gives confidence that the EBSD indexing has correctly distinguished between the pseudosymmetric $[001]_t$ and $[110]_t$ directions.

We note that in the center grain, there are a handful of points that were indexed as a different variant (magenta, such as the area labeled "g" in Fig. 8) compared to the rest of the grain (green). It is possible that these are misindexings. Looking at the experimental patterns at these points reveals that the quality of these patterns (Fig. 7g) is worse than in the rest of the grain (Fig. 7f). This could be a result of inadequate sample preparation or strains induced by grain boundaries or coherent monoclinic phase. If so, this would suggest that to resolve pseudosymmetry in $ZrO_2$-$13.5CeO_2$ for the most difficult orientations, we need a pattern quality similar to Fig. 7f. It is also possible that the indexing is correct, and a second variant is present that was formed during growth or induced by stress. Such cases have been reported in the literature (Virkar & Matsumoto, 1986; Jue & Virkar, 1990), although the sporadic locations and undefined morphology suggest that this may be unlikely.

*4.3. Pattern center sensitivity*

To confirm whether or not the indexed results shown in Fig. 7b are affected by small changes in pattern center, such as that between Eqs. (3) and (4), we varied the pattern center and reindexed selected patterns. Ten patterns were selected, one from each of the 10 largest grains (the same patterns used for the pattern center fitting). $X^*$ and $Y^*$ were varied by up to ±12px (2.5% detector width) while maintaining $Z^* = 15733$ μm. Because the optimal orientation is greatly altered by the large pattern center errors, up to ~3°, for each pattern center value the orientation of each pattern was first refined on a 7×7×7 grid with $w = 2.1°$. From this refined orientation, the patterns were then indexed on a 9×9×9 grid with $w = 2°$ followed by interpolation, as done throughout this paper.

Fig. 8a shows the fraction of the 10 patterns that were indexed to the correct pseudosymmetry variant as a function of deviation from the pattern center given in Eq. (4), assuming that the variant selected at zero error is correct. A relatively wide margin of error can be seen, as the smallest error that results in the first pattern being misindexed is ~6 px (1.25% detector width). Both the pattern centers found in Eqs. (3) and (4), which are indicated by the triangle and circle, respectively, are well within the bounds of 100% accuracy, giving confidence that the correct pseudosymmetry variant was selected. The minimum CI for the 10 patterns is shown in Fig. 8b, which reveals that the CI quickly approaches zero as the pattern center error is increased. Both the pattern centers from Eqs. (3) and (4) fall in the narrow region with high CI, giving further confidence in the fitted pattern centers.

In Fig. 8c, we have reindexed the dataset from Fig. 7b using an incorrect pattern center to illustrate the effect of pattern center errors on the resulting IPF map. As can be seen, a number of grains have been indexed as a different pseudosymmetry variant, as expected from the results in Fig. 8a. While the bottom-center grain clearly show two variants being selected, a single incorrect variant is selected in other grains (e.g. the center grain). Thus, it appears that dictionary indexing generally chooses one variant over another in a given grain/region, even if it is the incorrect variant and/or has low confidence in the selection, rather than exhibit a "checkboard" pattern as in Hough-based indexing. This makes it difficult to ascertain if pseudosymmetry has correctly been resolved in dictionary-indexed data solely by inspection of the IPF map. It is believed that this difference arises because of the error associated with the band detection step of Hough-based indexing; when the error in band detection is greater than the difference in band angles between different variants, the result is a seemingly random variant selection throughout the grain. Dictionary indexing, on the other hand, by directly comparing full pattern intensities eliminates this source of error and is thus more sensitive to small differences in patterns between pseudosymmetry variants. But its ability to select the correct variant is still limited by pattern center accuracy, noise, and other experimental factors that are not accurately reproduced in the simulations such as uncertainty in strain, lattice parameters, and optical distortions.

To rationalize these results, in Fig. 8d we have visualized the NDP function of the three pseudosymmetry variants for one of the patterns. It can be seen that the three surfaces do not necessarily trend together. Instead, they intersect, which defines regions of pattern center space in which that pattern is indexed as a given variant. The blue surface, corresponding to the correct variant, has the highest NDP around the fitted pattern center. But, in this example, the pattern would be misindexed with large positive errors in $X^*$ and $Y^*$ as the gray variant has a higher NDP in these regions of pattern center space. From analyzing the surfaces of all ten patterns (not shown), this domain for correct indexing varies for different orientations, and the intersection of these domains corresponds to Fig. 8a. This domain will likely shrink for materials with lower c/a ratios, but from these results, it appears that the presently proposed method can quite easily resolve pseudosymmetry in tetragonal $ZrO_2$-13.5$CeO_2$ as there is a rather large margin of error in pattern center to maintain 100% indexing accuracy as well as a large minimum CI when the correct pattern center is used for indexing.

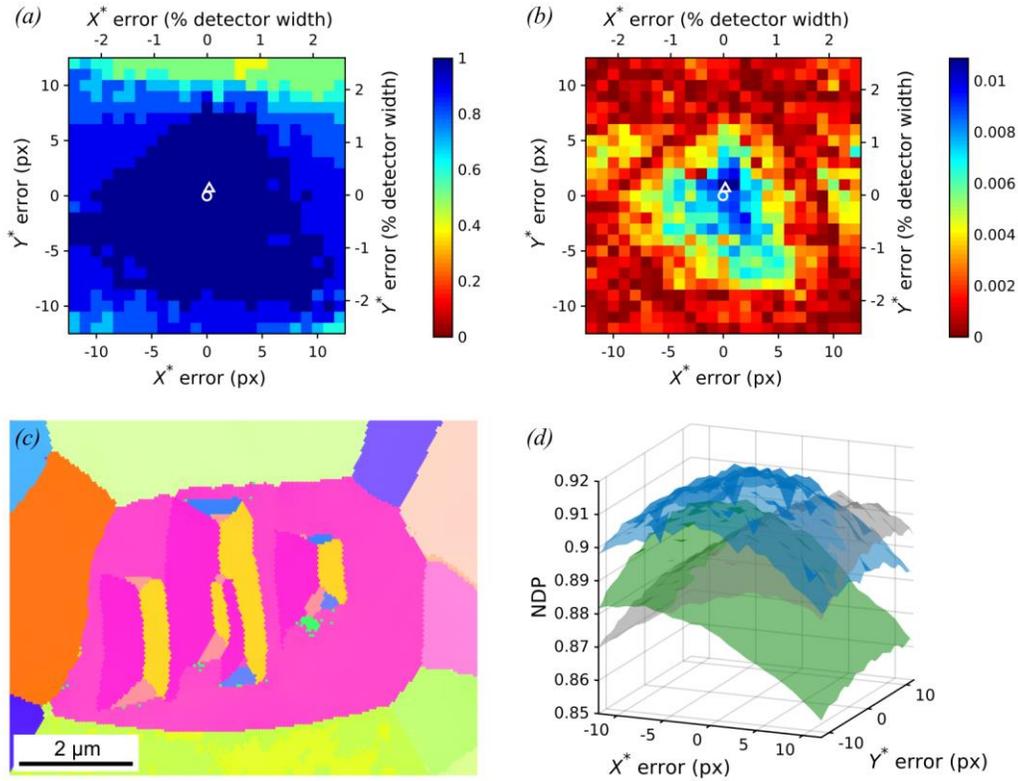

**Fig. 8.** Effects of pattern center errors on resolving pseudosymmetry in the experimental dataset from Fig. 7. (a) Fraction of patterns with the correct variant selected. Pattern centers from Eqs. (3) and (4) are labeled by a triangle and circle, respectively. (b) Minimum CI of the 10 patterns. (c) Reindexed map using an incorrect pattern center with errors in $X^*$ and $Y^*$ of -12 px. (d) NDP functions for the three pseudosymmetry variants for one of the patterns, illustrating that an incorrect variant (gray and green surfaces) can have a higher NDP than the correct variant (blue surface) at large pattern center errors.

## 5. Comparison to other methods for resolving pseudosymmetry

A number of approaches have been previously presented that aim to resolve pseudosymmetry with the aid of intensity-based indexing: 1) pattern matching, 2) cross-correlation EBSD (CC-EBSD), 3) dictionary indexing, and 4) spherical indexing. In this section, we compare these methods with the present method and highlight pros, cons, and opportunities for future improvements.

We note similarities of the present approach to the pattern matching method (Nolze *et al.*, 2016), which begins with Hough-indexed data and then fine-tunes the orientation of each map point using a downhill simplex algorithm, searching near the starting orientation as well as pseudosymmetric orientations. However, in our approach, we do not optimize each map point individually but rather feed in a dictionary of orientations in the spirit of dictionary indexing. This has two advantages: 1) speed, since we can take advantage of the rapid dot product computation on the GPU by *EMsoft* as opposed to many individual function calls during the optimization; and 2) eliminate misindexings, which may not be removed by pattern matching since it involves a local optimization procedure starting from the Hough-indexed orientation. Thus, in some ways, this method can be viewed as an efficient compromise between dictionary indexing and pattern matching.

CC-EBSD is another interesting approach to resolve pseudosymmetry that utilizes its strain-detecting capabilities. Jackson *et al.* (2018) demonstrated using simulated data that this method could correctly distinguish between pseudosymmetry variants of γ-TiAl with ~96% accuracy. In addition, their method could tolerate binning down to a size of 82×82 and pattern center errors of up to 0.054% detector width. In similar tests on simulated data, our method achieves 100% accuracy down to a pattern size of 60×60 in the presence of significant amounts of noise (data in Fig. 6, representative pattern shown in Fig. 2) and can also tolerate pattern center errors of up to ~1.25% detector width. While it is difficult to directly compare performance (since we studied tetragonal $ZrO_2$ with c/a = 1.0185 as opposed to γ-TiAl with c/a = 1.016), it is believed that the present method has at least comparable performance to the CC-EBSD approach.

De Graef *et al.* (2019) used a strategy involving dictionary indexing plus an additional orientation refinement step to resolve pseudosymmetry in $CuInSe_2$ with c/a = 1.003, the lowest c/a ratio attempted to date for the resolution of pseudosymmetry by EBSD. Theirs is similar to the method used in the present study, but the primary differences are: 1) we use a global search method with multiple patterns using the *pcglobal* package to identify the pattern center with high accuracy, rather than a single high-quality pattern using the built-in *EMsoft* routines; 2) we use a focused dictionary based on the Hough-indexed data rather than a full dictionary covering all of orientation space for greater computational speed and efficiency; and 3) we refine the orientations by interpolation as opposed to using a local optimization algorithm. We note that optimizing the pattern center and orientations by local search algorithms such as Nelder-Mead simplex or BOBYQA, which are used in the pattern center fitting and orientation optimizing routines of *EMsoft*, can easily get stuck in local maxima (Tanaka & Wilkinson, 2019; Pang *et al.*, 2020; Rios & Sahinidis, 2013), which can result in very large pattern center errors (Fig. 3) that compromise the ability of the indexing to correctly resolve pseudosymmetry (Fig. 8). However, while their method removed most of the ambiguous indexings from the Hough data, the present results demonstrate that umambigious indexing is an inherent property of dictionary indexing, and removal of ambiguous indexings does not necessarily mean that the patterns have been correctly indexed (Fig. 8c). It has not yet been conclusively demonstrated that their method, or any method to date, can correctly resolve pseudosymmetry in materials with c/a ratios as low as 1.003. Testing of our method on materials with such low c/a ratios is currently ongoing, and the results will be presented in a forthcoming publication.

The recent spherical indexing method has also been applied to resolve pseudosymmetry (Lenthe *et al.*, 2019*b*). The key parameter in this approach is the bandwidth, where smaller bandwidth values result in more significant approximations but faster indexing, in some cases as fast as Hough-based indexing. In a pseudosymmetric forsterite material (Lenthe *et al.*, 2019*b*), spherical indexing with a bandwidth of 68 resulted in a "checkerboard" pattern, similar to Hough-based indexing, whereas using a larger bandwidth of 263 appeared to resolve pseudosymmetry. However, resolving pseudosymmetry in forsterite is considerably easier than in tetragonal $ZrO_2$, and it remains to be seen how accurately spherical indexing can distinguish between pseudosymmetry variants when there is no discernable difference in their patterns to the human eye (Fig. 1). In such challenging materials, even larger bandwidths would likely be required, at which point it may be computationally more expensive than dictionary indexing (Lenthe *et al.*, 2019*a*).

The present results give us optimism that a combination of approaches that include our methods here and elements of those above will lead to a general path for EBSD indexing of even the most subtle situations involving pseudosymmetry. For example, one challenge for the present method would be if there are no correctly indexed points within a grain and no correctly indexed grains with similar orientation; in such a case, our method will not be able to correctly index that grain, although this appears rather unlikely for most datasets. In such a case, it may be better to first perform a standard dictionary indexing or spherical indexing run and then refine the data using the present methodology. The present algorithm is most efficient when the input data is of reasonably good quality, the number of grains in the map is low, and when the primary objective for using an advanced EBSD analysis method is to resolve pseudosymmetry.

## 6. Conclusions

In this work, we have outlined a method for resolving pseudosymmetry in EBSD using a modified dictionary indexing approach, where we incorporate accurate pattern center fitting using global optimization, clustering of the Hough-indexed data to focus the dictionary in orientation space, and interpolation to improve the accuracy of the indexed solution (summarized in Fig. 5 and Section 2.4). On a simulated dataset of $ZrO_2$-13.5$CeO_2$, with $c/a = 1.0185$, our method selects the correct pseudosymmetry variant for all 100 patterns tested over all regions of orientation space, even for patterns with 8×8 binning (60×60 px image size) and considerable noise. We then applied the present method to an experimental dataset of $ZrO_2$-13.5$CeO_2$ containing both the pseudosymmetric tetragonal and non-pseudosymmetric monoclinic phases. The ambiguous indexing of the tetragonal phase from Hough-based indexing was fully resolved by the present method. In addition, the observed lattice correspondence ($[010]_m \parallel [001]_t$) between the tetragonal and monoclinic phases is in agreement with expectation, giving confidence that our indexing method has correctly resolved pseudosymmetry. We find that fitting an accurate pattern center is crucial to correctly resolving pseudosymmetry, as an inaccurate pattern center can lead to an unambiguous but incorrect indexing. This indexing method only took 45 min to index the tetragonal phase on our modest 8-core workstation, compared to 2.7 hours for dictionary indexing with typical parameters, which is unable to resolve pseudosymmetry in this material without an additional refinement procedure. Based on the relatively high CI values observed, our method can likely be applied to even more challenging pseudosymmetric materials than $ZrO_2$-13.5$CeO_2$ in the future. Our code implementation is available online at: https://github.com/epang22/EBSDrefine.

**Acknowledgements**

The authors would like to thank Prof. Marc De Graef (Carnegie Mellon University) for providing scripts to interface with *EMsoft* and assisting with the dictionary indexing method.

**Funding information**

This material is based upon work sponsored in part by the U.S. Army Research Office through the Institute for Soldier Nanotechnologies, under Cooperative Agreement number W911NF-18-2-0048. This work made use of the MRSEC Shared Experimental Facilities at MIT, supported by the NSF under award number DMR-1419807. Financial support is also acknowledged from the NSF Graduate Research Fellowship Program under grant number DGE-1745302 (E.L.P.) and the Danish Council for Independent Research under grant number 7026-00126B (P.M.L.).

**References**

Brewer, L. N., Kotula, P. G. & Michael, J. R. (2008). *Ultramicroscopy*. **108**, 567–578.
Brough, I., Bate, P. S. & Humphreys, F. J. (2006). *Mater. Sci. Technol.* **22**, 1279–1286.
Chen, Y. H., Park, S. U., Wei, D., Newstadt, G., Jackson, M. A., Simmons, J. P., De Graef, M. & Hero, A. O. (2015). *Microsc. Microanal.* **21**, 739–752.
Demirel, M. C., El-Dasher, B. S., Adams, B. L. & Rollett, A. D. (2000). *Electron Backscatter Diffraction in Materials Science*, Vol. edited by A.J. Schwartz, M. Kumar & B.L. Adams, pp. 65–74. New York: Springer.
Engler, O. & Randle, V. (2010). Introduction to texture analysis: macrotexture, microtexture and orientation mapping Boca Raton: CRC Press.
Foden, A., Collins, D. M., Wilkinson, A. J. & Britton, T. B. (2019). *Ultramicroscopy*. **207**, 112845.
De Graef, M., Lenthe, W. C., Schäfer, N., Rissom, T. & Abou-Ras, D. (2019). *Phys. Status Solidi RRL*. **13**, 1900032.
Humphreys, F. J. (1999). *J. Microsc.* **195**, 170–185.
Humphreys, F. J. (2001). *J. Mater. Sci.* **36**, 3833–3854.
Huyer, W. & Neumaier, A. (2008). *ACM Trans. Math. Softw.* **35**, 1–25.

Jackson, B. E., Christensen, J. J., Singh, S., De Graef, M., Fullwood, D. T., Homer, E. R. & Wagoner, R. H. (2016). *Microsc. Microanal.* **22**, 789–802.

Jackson, B., Fullwood, D., Christensen, J. & Wright, S. (2018). *J. Appl. Crystallogr.* **51**, 665–669.

Jackson, M. A., Pascal, E. & De Graef, M. (2019). *Integr. Mater. Manuf. Innov.* **8**, 226–246.

Jimenez, R., Bucheli, W., Varez, A. & Sanz, J. (2011). *Fuel Cells*. **11**, 642–653.

Jue, J. F. & Virkar, A. V. (1990). *J. Am. Ceram. Soc.* **73**, 3650–3657.

Kelly, P. M. & Francis Rose, L. R. (2002). *Prog. Mater. Sci.* **47**, 463–557.

Krieger Lassen, N. C. (1994). Automated Determination of Crystal Orientations from Electron Backscattering Patterns. Technical University of Denmark.

Lenthe, W. C., Singh, S. & De Graef, M. (2019*a*). *Ultramicroscopy*. **207**, 112841.

Lenthe, W., Singh, S. & De Graef, M. (2019*b*). *J. Appl. Crystallogr.* **52**, 1157–1168.

Marquardt, K., De Graef, M., Singh, S., Marquardt, H., Rosenthal, A. & Koizuimi, S. (2017). *Am. Mineral.* **102**, 1843–1855.

Martin, S., Berek, H., Aneziris, C. G., Martin, U. & Rafaja, D. (2012). *J. Appl. Crystallogr.* **45**, 1136–1144.

Nolze, G., Hielscher, R. & Winkelmann, A. (2017). *Cryst. Res. Technol.* **52**, 1600252.

Nolze, G., Winkelmann, A. & Boyle, A. P. (2016). *Ultramicroscopy*. **160**, 146–154.

Nowell, M. M. & Wright, S. I. (2005). *Ultramicroscopy*. **103**, 41–58.

Ocelík, V., Schepke, U., Rasoul, H. H., Cune, M. S. & De Hosson, J. T. M. (2017). *J. Mater. Sci. Mater. Med.* **28**, 121.

Pang, E. L., Larsen, P. M. & Schuh, C. A. (2020). *Ultramicroscopy*. **209**, 112876.

Pang, E. L., McCandler, C. A. & Schuh, C. A. (2019). *Acta Mater.* **177**, 230–239.

Pee, J. H., Tada, M. & Hayakawa, M. (2006). *Mater. Sci. Eng. A*. **438–440**, 379–382.

Ram, F., Wright, S., Singh, S. & De Graef, M. (2017). *Ultramicroscopy*. **181**, 17–26.

Rios, L. M. & Sahinidis, N. V. (2013). *J. Glob. Optim.* **56**, 1247–1293.

Ryde, L. (2006). *Mater. Sci. Technol.* **22**, 1297–1306.

Singh, S. & De Graef, M. (2016). *Model. Simul. Mater. Sci. Eng.* **24**, 085013.

Singh, S., Ram, F. & De Graef, M. (2017). *J. Appl. Crystallogr.* **50**, 1664–1676.

Tanaka, T. & Wilkinson, A. J. (2019). *Ultramicroscopy*. **202**, 87–99.

Venables, J. A. & Harland, C. J. (1973). *Philos. Mag.* **27**, 1193–1200.

Virkar, A. V. & Matsumoto, R. L. K. (1986). *J. Am. Ceram. Soc.* **69**, C224–C226.

Wilkinson, A. J., Collins, D. M., Zayachuk, Y., Korla, R. & Vilalta-Clemente, A. (2019). *Ultramicroscopy*. **196**, 88–98.

Wilkinson, A. J., Dingley, D. J. & Meaden, G. (2009). *Electron Backscatter Diffraction in Materials Science*, Vol. edited by A.J. Schwartz, M. Kumar, B.L. Adams & D.P. Field, pp. 231–249. New York: Springer.

Wright, S. I. (1992). Individual lattice orientation measurements: Development and application of a fully automated technique. Yale University.

Wright, S. I. (2000). *Electron Backscatter Diffraction in Materials Science*, Vol. edited by A.J. Schwartz, M. Kumar & B.L. Adams, pp. 51–64. New York: Springer.

Wright, S. I. (2006). *Mater. Sci. Technol.* **22**, 1287–1296.

Wright, S. & Nowell, M. (2008). *Adv. Mater. Process.* **166**, 29–32.

Yashima, M., Hirose, T., Katano, S., Suzuki, Y., Kakihana, M. & Yoshimura, M. (1995). *Phys. Rev. B*. **51**, 8018–8025.

Zambaldi, C., Zaefferer, S. & Wright, S. I. (2009). *J. Appl. Crystallogr.* **42**, 1092–1101.